%% file: example_paper.tex
\documentclass{article}

\usepackage{microtype}
\usepackage{graphicx}
\usepackage{subfigure}
\usepackage{booktabs} % for professional tables

\usepackage{hyperref}
\usepackage{xurl}
\usepackage{xspace}
\usepackage{enumitem}
\usepackage{placeins} % provides \FloatBarrier

\usepackage[accepted]{mlsys2025}
\makeatletter
\renewcommand{\Notice@String}{}
\renewcommand{\printAffiliationsAndNotice}[1]{%
\stepcounter{@affiliationcounter}%
{\let\thefootnote\relax\footnotetext{\hspace*{-\footnotesep}#1%
\forloop{@affilnum}{1}{\value{@affilnum} < \value{@affiliationcounter}}{%
\textsuperscript{\arabic{@affilnum}}\csname @affilname\the@affilnum\endcsname}.}}}
\def\bottomtitlebar{\vskip .15in \hrule height1pt \vskip .2in}
\makeatother
\hypersetup{pdfsubject={}}
\newcommand{\sysname}{\textsc{CoEvoAgent}\xspace}
\newcommand{\bench}{\textsc{CoEvoBench}\xspace}
\providecommand{\Description}[1]{}

\mlsystitlerunning{Keeping Up with the Kernel}

\begin{document}

\twocolumn[
\mlsystitle{Keeping Up with the Kernel: Can LLMs Adapt Linux Drivers to Interface Changes?}

\begin{mlsysauthorlist}
\mlsysauthor{Xinrui Yang}{mbzuai}
\mlsysauthor{Jiawen Liu}{mbzuai}
\mlsysauthor{Arina Kharlamova}{mbzuai}
\mlsysauthor{Humaid Alqasimi}{mbzuai}
\mlsysauthor{Tianyi Zhang}{mbzuai}
\mlsysauthor{Alnoor Ismail}{mbzuai}
\mlsysauthor{Youcheng Sun}{mbzuai}
\mlsysauthor{Chun Jason Xue}{mbzuai}
\end{mlsysauthorlist}

\mlsysaffiliation{mbzuai}{Mohamed bin Zayed University of Artificial Intelligence, Abu Dhabi, United Arab Emirates}

\mlsyskeywords{Linux Device Drivers, Collateral Evolution, Benchmark, Large Language Models}

\vskip 0.15in
\begin{abstract}
\input{appendix/driverbench/abstract}
\end{abstract}
]

\printAffiliationsAndNotice{}

\input{appendix/driverbench/sections/introduction}
\input{appendix/driverbench/sections/background}
\input{appendix/driverbench/sections/benchmark}
\input{appendix/driverbench/sections/pipeline}
\input{appendix/driverbench/sections/experiments}
\input{appendix/driverbench/sections/discussion}
\input{appendix/driverbench/sections/related_work}
\input{appendix/driverbench/sections/conclusion}

% Slightly tighter spacing between reference entries (template: 8pt) so
% the reference list ends one page earlier. Layout only.
\setlength{\bibsep}{4pt plus 1pt minus 1pt}
\bibliography{references}
\bibliographystyle{mlsys2025}

\clearpage
\section*{AI Usage Disclosure}

The large language models named in Section~\ref{sec:experiments-models}
(Claude Sonnet 5.5, GPT-6.1 Sol, and DeepSeek) are objects of study in this
work. Their outputs are reported only as part of the experimental analysis.
Separately, the authors used Claude (Anthropic) as an assistant for parts of
the engineering and writing. This assistance covered helping the authors
understand Linux kernel data, drafting and reviewing data processing and plotting scripts,
checking bibliographic entries against
the original sources, and editing the English of manuscript drafts. The
research ideas and the core design of this work did not originate from AI
tools. All research questions, benchmark design, task selection criteria, experimental
protocols, and conclusions were defined by the authors. The authors reviewed
and verified every script, number, citation, and passage that AI tools helped
produce, and they take full responsibility for the content of this paper.
No AI tool is listed as an author.

\end{document}

%% file: appendix/driverbench/abstract.tex
We analyze \emph{collateral evolution} in Linux, the coordinated updates that
drivers require when the kernel interfaces they depend on change, over the 20
years since it was first studied, covering releases from v3.0 to v7.2. Since
v2.6.13, the number of driver files and support libraries has grown about
eightfold, a single change to the most widely used library can now reach
about 5.3k driver files, and such changes arrive either in bursts or spread
over many releases.
We introduce \bench, a benchmark that asks models to adapt an old driver
module to a target kernel and evaluates the result against the developer's
reference patch. The benchmark contains 316 manually confirmed tasks from
202 mainline commits and 268 modules, spanning kernel releases v6.19 to v7.2.
It distinguishes 134 \emph{build-breaking} tasks, where the old driver fails
to compile, from 182 \emph{build-compatible} tasks, where the old usage
still builds. Using \sysname{}, we evaluate Direct, SmPL, and mini-SWE-agent
with DeepSeek Flash, and compare Direct with Claude Sonnet 5.5 and GPT-6.1
Sol. With DeepSeek, default Direct generation achieves strict reference
agreement on 41.0\% of build-breaking tasks but only 3.3\% of
build-compatible tasks.

%% file: appendix/driverbench/sections/introduction.tex
\section{Introduction}
\label{sec:introduction}

Linux device drivers depend on interfaces provided by the rest of the kernel,
such as library functions, callbacks, data structures, and
protocols~\cite{padioleau2006collateral}. The kernel does not keep these
interfaces stable~\cite{kh_stableapi}, so changes can require updates to the
drivers that use them. Padioleau et al.\ call such updates \emph{collateral
evolutions}~\cite{padioleau2006collateral}. In mainline Linux, an old interface
is typically removed after its in-tree users have been converted, while
out-of-tree code must keep up with new kernels on its
own~\cite{cook2018ksummit,corbet2018deprecated,kernel_process_intro}. Vendors
may therefore need to port drivers as their customers move to newer
kernels~\cite{rodriguez2015backports}. Our study of driver evolution since v3.0
(Section~\ref{sec:driver-evolution}) shows that widely used support libraries
now reach more driver files, and at least half of the function replacements we
examine span multiple release updates.

Following an interface change correctly requires understanding how each driver
uses it. This usage context can vary across device-specific
files~\cite{padioleau2006collateral}, so a rule that works for one driver may
miss an important condition in another. Coccinelle, a C matching and
transformation tool widely used by kernel developers, applies rules written in
the Semantic Patch Language (SmPL) and was the basis of over 6000 kernel
commits as of Linux 4.15~\cite{lawall2018coccinelle}. One semantic patch
converted seven LED drivers to the managed interface
\texttt{devm\_led\_classdev\_register}~\cite{led_devm_conversions}. Managed
resources are released when their associated device is
detached~\cite{kernel_devres}. Four of these drivers registered the LED under
the parent device, however, so registration outlived memory freed when the
child driver was unbound. The resulting use-after-free remained in mainline for
more than four years before all four drivers were fixed~\cite{led_fixes}. The
rule matched the code but did not distinguish the relevant device lifetimes;
compilation alone could not expose the bug. This illustrates why a migration
must account for driver context as well as interface usage.

This raises the question of whether LLMs, increasingly used for code
generation and program repair~\cite{jimenez2024swebench,yang2024sweagent}, can
adapt drivers to interface changes in their own contexts. Existing work
addresses related but distinct problems. Prequel retrieves historical commits
for errors encountered when porting drivers to a target kernel, leaving the
changes to the developer; its authors also observe that a compiler need not
report every required change~\cite{lawall2017prequel}. MigGPT migrates
out-of-tree kernel patches across versions~\cite{miggpt2025}. kGym and
Live-kBench evaluate LLMs on kernel crash resolution with
reproducers~\cite{mathai2024kgym,huang2026livekbench}. To our knowledge, no
benchmark evaluates whether LLMs can update Linux driver modules in response
to real interface changes against the changes made by kernel developers.

We present \bench, a benchmark for evaluating collateral evolutions in Linux
drivers. Its tasks are derived from public mainline commits, where developer
migrations provide reviewed references for assessing generated changes. Each
task supplies the pre-change source of a driver module and a target kernel,
and asks a method to update the module. We rebuild the old driver against the
target kernel and distinguish \emph{build-breaking} tasks, where compilation
fails, from \emph{build-compatible} tasks, where it succeeds. For the latter,
the original usage still builds even though developers adopted a recommended
change. The distinction matters for evaluation: a successful build can reveal
that a patch resolves a compilation failure, but it cannot establish whether
a recommended migration has been carried out. Reference implementations
provide another point of comparison, while allowing for the possibility that
different source changes implement the same adaptation.
\bench contains 316 tasks from 202 commits and 268 modules across the
three release updates from v6.19 to v7.2. The tasks include 134 build-breaking
and 182 build-compatible cases, grouped by 91 interface changes; we manually
confirmed the relationship between each task and its interface change.

Using \sysname{} with DeepSeek Flash, we evaluate direct patch generation,
SmPL generation executed by Coccinelle, and mini-SWE-agent. We also compare
direct generation with Claude Sonnet 5.5 and GPT-6.1 Sol, reporting compilation
and strict reference agreement separately. Default Direct generation with
DeepSeek reaches 41.0\% strict agreement on build-breaking tasks but 3.3\% on
build-compatible tasks. Annotated old-API localization hints raise agreement on
the latter to 10.4\%, showing the importance of identifying the applicable
migration when the old usage still builds. To interpret reference differences,
we also review compile-passing DeepSeek Direct patches whose source is not
identical to the developer version. The review finds both acceptable
alternative implementations and patches with defects or incomplete migrations,
making the distinction between build success and migration quality concrete.

This paper makes the following contributions:
\begin{itemize}
  \item \textbf{An analysis of Linux driver interface evolution from v3.0 to
    v7.2}, measuring driver dependencies on support libraries and the release
    updates spanned by function replacements.
  \item \textbf{\bench}, a benchmark of 316 manually confirmed driver
    adaptation tasks from 91 interface changes, accompanied by \sysname{} and
    the validation scripts used in our evaluation.
  \item \textbf{An evaluation of adaptation strategies and models} that
    examines the effects of migration localization and compiler feedback, and
    reports buildability and reference agreement separately.
\end{itemize}

Code and data are available online.\footnote{\url{https://anonymous.4open.science/r/3C84/}}

%% file: appendix/driverbench/sections/background.tex
\section{Revisiting Collateral Evolution Twenty Years Later}
\label{sec:driver-evolution}

This section examines how Linux drivers have followed kernel interface changes
from v3.0 to v7.2, measuring how widely driver files depend on support
libraries (Section~\ref{sec:driver-evolution-reach}) and how collateral
evolutions are distributed over release updates
(Section~\ref{sec:driver-evolution-repeated}), and derives from these
observations the design of \bench{}
(Section~\ref{sec:driver-evolution-implications}).

Drawing on the study of Padioleau et al.\ on Linux v2.2.0 to
v2.6.13~\cite{padioleau2006collateral}, whose figures we cite for those
versions, we separate support libraries from driver code. Within each
directory, the files that export functions to multiple other files form the
support library, which provides common services for one class of devices, and
the remaining files are driver code. We measure collateral evolutions by the
number of lines of driver code they modify in each release update, from one
release to the next (e.g., v4.16 to v4.17).

\subsection{A single interface change now reaches more drivers}
\label{sec:driver-evolution-reach}

\begin{figure*}[t]
  \centering
  \includegraphics[width=\textwidth]{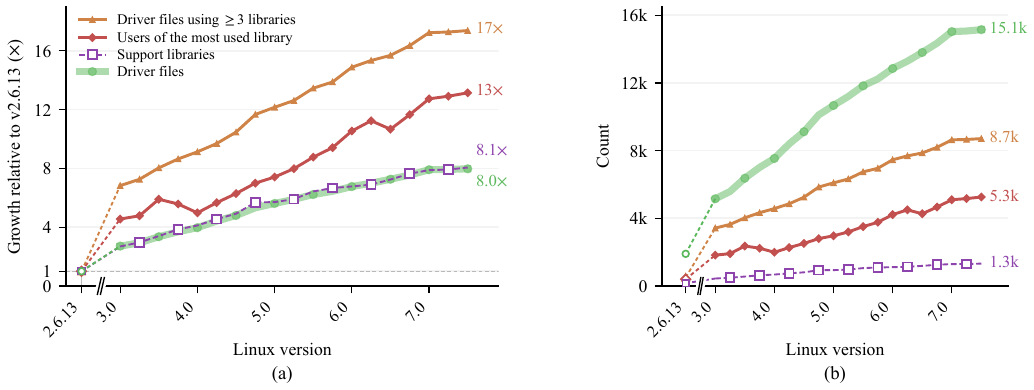}
  \vspace{-8pt}% template adds 10pt above every caption
  \caption{Driver dependencies since the 2006 study, across 19 Linux
  versions: the .0, .5, .10, and .15 releases of each series from v3.0 to
  v6.15, plus v7.0, v7.1, and v7.2. The leftmost points are the counts
  Padioleau et al.\ reported for v2.6.13: 1.9k driver files, 164 support
  libraries, about 400 driver files using the most widely used library (PCI),
  and 501 driver files using three or more support libraries.
  (a) Growth of each count relative to v2.6.13. From v3.0 onward, the most
  widely used library is \texttt{drivers/base}.
  (b) The same four counts as absolute values.}
  \Description{Two panels. The left panel shows four lines rising from the 2006 baseline of one to about eight, eight, thirteen, and seventeen times by v7.2. The right panel shows the same quantities as absolute counts on a linear axis, reaching 15.1k driver files, 1.3k libraries, 5.3k files using the most widely used library, and 8.7k files using three or more libraries in v7.2.}
  \label{fig:followup-dependencies}
\end{figure*}

Two quantities describe how drivers depend on support libraries: how many
interfaces each driver file uses, which determines how exposed a single file is
to interface changes, and how many driver files use each library, which
determines how many files a single change in that library reaches.
Figure~\ref{fig:followup-dependencies} compares the resulting counts across
19 versions with those reported for v2.6.13.

Across these versions, the median number of interfaces per driver file
changes little. The median number of interface
functions referenced by a driver file fell slightly from 21 in v3.0 to 19 in
v7.2, and the P90 stayed between 52 and 57. The P99 rose from 108 in v3.0 to
about 120 by v4.0 and has stayed between 121 and 124 since. Files that depend
on three or more support libraries (Figure~\ref{fig:followup-dependencies})
made up 66\% of driver files in v3.0 and about 57\% from v4.5 onward, while
their number grew from 3.4k to 8.7k, about 17 times the 501 reported for
v2.6.13.

What has grown is mainly the number of driver files that share a widely used
library. Padioleau et al.\ reported 1.9k driver files and 164 support
libraries in v2.6.13. We identify 5.15k and 441 in v3.0, and 15.1k and 1.3k in
v7.2, about eight times the 2006 counts. The median library is used by 4
driver files in every version. The
most widely used library, \texttt{drivers/base}, is used by about 35\% of
driver files in both v3.0 and v7.2, so as the driver population grew, its
users nearly tripled, from 1.8k to 5.3k. In v2.6.13, the most widely used
library, PCI, was used by roughly one fifth of driver files, about 400.

\textbf{A single interface change in the most widely used library can thus
reach about 5.3k driver files in v7.2, more than ten times the roughly 400
that PCI reached in v2.6.13.}

\subsection{Collateral evolutions arrive both in bursts and over many releases}
\label{sec:driver-evolution-repeated}

\begin{figure*}[t]
  \centering
  \includegraphics[width=\textwidth]{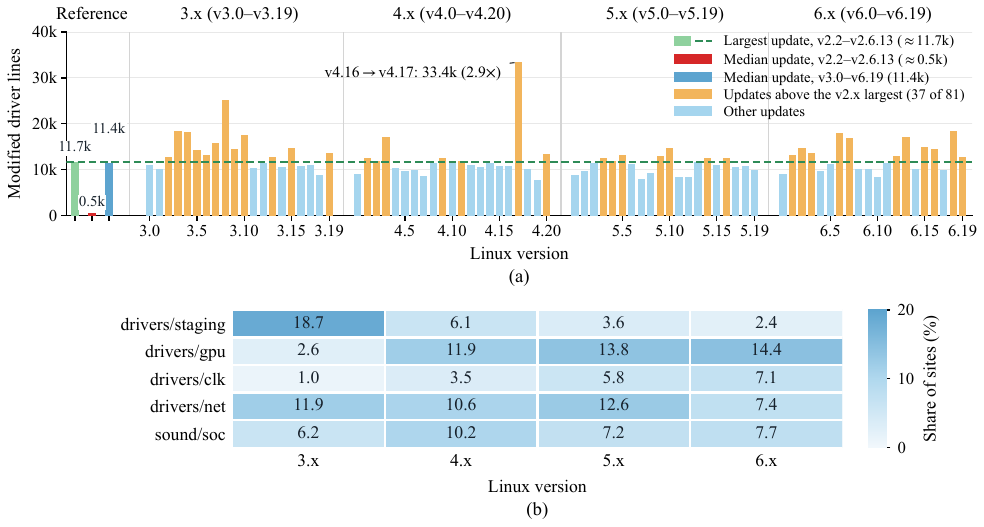}
  \vspace{-8pt}% template adds 10pt above every caption
  \caption{(a) Lines of driver code modified by collateral evolutions in each
  of the 81 consecutive release updates into v3.0 through v6.19, grouped by the
  Linux 3.x, 4.x, 5.x, and 6.x series. The three reference bars on the left
  give the largest update (about 11.7k lines) and the median update (about
  0.5k) of v2.2 to v2.6.13 in the 2006 study, measured from its Figure~14,
  and the median of our 81 updates (11.4k). The dashed line marks 11.7k, and
  the 37 orange bars exceed it.
  (b) Share of all collateral-evolution sites in each series that fall in each
  of five directories.}
  \Description{The upper panel shows, for eighty-one release updates in four separated groups, the number of driver code lines modified by collateral evolutions. The lower heatmap shows the percentage of collateral-evolution sites in staging, GPU, clk, net, and sound/soc within each of the four series. The staging share decreases and the GPU share increases.}
  \label{fig:followup-repeated-edits}
\end{figure*}

Collateral evolutions reach drivers mainly in two forms: concentrated
migrations completed within a single release update, and gradual migrations
spread over many updates. Figure~\ref{fig:followup-repeated-edits} covers the
complete Linux 3.x to 6.x series, from the first update of 3.x (v2.6.39 to
v3.0) to the last release of 6.x (v6.19). The 7.x series, still in progress,
is excluded. \textbf{A release update modifies a median of about 11.4k lines
of driver code through collateral evolutions
(Figure~\ref{fig:followup-repeated-edits}(a)), close to the highest value in
the 2006 study (about 11.7k, measured from its Figure~14, where the median
update from v2.2.1 to v2.6.13 is about 0.5k), and the largest update, v4.16
to v4.17, reaches about 33.4k, nearly three times that peak.}
Padioleau et al.\ observed the highest shares of such lines (up to 35\%) in the
unstable development series 2.3 and 2.5. Since Linux abandoned separate
development series in 2004, peaks now occur between regular releases.

Nine of the 81 updates exceed the median by more than half. We manually
examined the six highest. In four, the collateral evolution with the most
sites was submitted mainly by one or two developers within a single release
cycle. In v4.16 to v4.17, where 60\% of the sites fall under
\texttt{sound/soc}, one ASoC developer carried out most of the replacement of
the \texttt{codec} object with \texttt{component} across the sound-card
drivers, in nearly 200 commits. The other three are the removal of obsolete
annotations such as \texttt{\_\_devinit} from 2.7k files by two developers
within five weeks (v3.7 to v3.8), the move of DVB front ends to the v5
interface (v3.2 to v3.3), and the ALSA conversion of lock handling to
\texttt{guard()} (v6.17 to v6.18).

These migrations were triggered by interface changes in support libraries, by
changes in core kernel conventions, and by style cleanups within a subsystem.
Figure~\ref{fig:followup-repeated-edits}(b) also shows that the subsystems
where collateral evolutions concentrate shift over time: the share of sites
under \texttt{drivers/staging} fell from 19\% to 2\%, whereas that under
\texttt{drivers/gpu} rose from 3\% to 14\%.

Many other collateral evolutions are completed gradually. We consider those of
the form ``old function replaced by new function'' with different old and new
names, the kind closest to an interface replacement. Of the 4.3k such
collateral evolutions, counted separately in each series, more than half
appear in multiple release updates rather than within one. For those spanning
multiple updates, the median number of updates from first to last appearance
is 9 to 12 within each of the 3.x, 4.x, 5.x, and 6.x series, about one and a
half to two years.

Replacements that touch more subsystems are generally less likely to finish
within one update. We approximate subsystems by two-level directories such as
\texttt{drivers/net} or \texttt{sound/soc}. Among replacements with at least
20 sites, 70\% of the 397 confined to one directory finish within one update,
compared with about 10\% of the 175 touching ten or more directories, roughly
a sevenfold gap.

We examined ten cross-subsystem migrations in the git history and found that
their replacement commits mainly follow two patterns:
\begin{itemize}
  \item One or a few developers advance the migration subsystem by subsystem
  over several years. The change of the platform-driver \texttt{remove}
  callback's return type, for example, was carried out mainly by the developer
  who introduced the new callback, in about 2k commits.
  \item Many developers contribute in a distributed fashion. In the
  replacement of \texttt{strlcpy} by \texttt{strscpy}, the new interface was
  added in 2015, and by the time the old interface was removed in v6.8, 49
  developers had submitted 159 replacement commits.
\end{itemize}

\subsection{Implications for the benchmark}
\label{sec:driver-evolution-implications}

Based on these observations, we design \bench{} as follows. A single
interface change in a widely used library can reach up to about 5.3k driver files
(Section~\ref{sec:driver-evolution-reach}), so \bench{} groups tasks by the
interface change that triggers them and caps the number of tasks per group to
keep any single change from dominating the benchmark
(Section~\ref{sec:benchmark-construction}). Since more than half of function replacements span multiple updates (Section~\ref{sec:driver-evolution-repeated}), \bench{} takes
the change of one commit to one driver module as the unit of a task. Its three
consecutive release updates, from v6.19 to v7.2, include both drivers updated
together with an interface change and drivers updated several releases later.

%% file: appendix/driverbench/sections/benchmark.tex
\section{\bench}
\label{sec:driver-bench}

This section defines the tasks of \bench{} (Section~\ref{sec:benchmark-task}),
describes how they are built from three consecutive Linux release updates and
confirmed manually (Section~\ref{sec:benchmark-construction}), and summarizes
their coverage, interface kinds, and size
(Section~\ref{sec:benchmark-features}).

\subsection{Task Formulation}
\label{sec:benchmark-task}

\paragraph{Interfaces.}
An interface that a driver depends on mainly comprises four kinds of
elements~\cite{padioleau2006collateral}: library functions, which the kernel
or a support library exports; callbacks, which a driver implements for the
library to call; data structures; and protocols, such as call ordering,
locking, and error checking. When an interface changes (e.g., an old usage is
removed or deprecated, or a new one is recommended), the drivers that use it
are expected to change accordingly, and each such driver-side change is part
of a \emph{collateral evolution}.

\paragraph{Build-breaking and build-compatible tasks.}
We classify each task by reverting the driver change and rebuilding the module
against the target kernel. When a developer makes a backward-incompatible
change to an interface and updates every in-tree user in the same
change~\cite{kh_stableapi}, as the patch-submission guidelines encourage by
asking that each patch keep the kernel buildable~\cite{kernel_submitting_patches},
reverting the driver part of that change leaves the module using an interface
that no longer exists. The rebuild then fails, and we call the task
\emph{build-breaking}. Other changes leave the old usage in place. A
deprecated interface cannot be replaced everywhere within one development
cycle, so the old and new usages coexist while drivers
migrate~\cite{kernel_deprecated}. Such migrations can last for years. The
migration from \texttt{init\_timer} to \texttt{setup\_timer}, for example, ran
from 2006 to 2017~\cite{serrano2020spinfer}, and once \texttt{timer\_setup} was
introduced in 2017, calls to \texttt{setup\_timer} could in turn be replaced at
leisure~\cite{corbet2017timers}. Reverting a driver change of this kind still
builds, and we call the task \emph{build-compatible}. \bench{} contains tasks
of both kinds (Section~\ref{sec:benchmark-construction}).

\begin{figure}[t]
  \centering
  \includegraphics[width=0.9\linewidth]{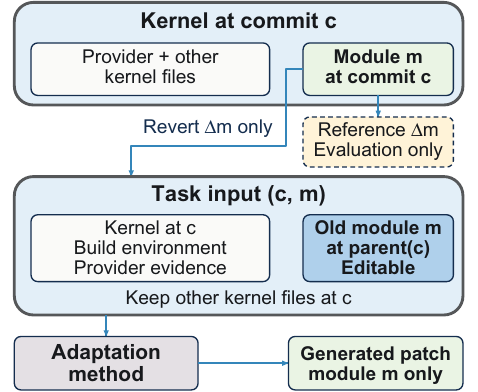}
  \caption{Task formulation. A task $(c, m)$ asks a method to adapt module
    $m$ from the parent of commit $c$ to the interfaces at
    $c$, modifying only the files of $m$. $\Delta_m$ serves as the reference
    answer.}
  \label{fig:task-formulation}
  \Description{Top: the kernel at commit c, with the provider and other
    kernel files and module m; reverting Delta m gives the task input, and
    Delta m is kept as the reference answer for evaluation only. Middle: the
    task input, with the kernel at c and the old module m at the parent of c,
    which is editable. Bottom: an adaptation method produces a patch that
    changes only module m.}
\end{figure}

\paragraph{Task unit.}
\bench organizes the changes drivers make in response to interface changes
into tasks. A driver module is one Kbuild target, that is, the source files
and local headers built into one \texttt{.ko}. We use \emph{method} for any
approach that produces a patch for a task. The change of a commit $c$ can be
split into three parts:
\[
c = \Delta_{\mathrm{provider}} + \sum_{m} \Delta_m + \Delta_{\mathrm{other}},
\]
where $\Delta_{\mathrm{provider}}$ is the modification of the interface
provider, $\Delta_m$ is the modification of a driver module $m$ (the sum runs
over the modules $c$ touches), and $\Delta_{\mathrm{other}}$ comprises the remaining
changes, such as documentation or non-driver code. The cause of a driver's change
is not necessarily in $c$: it may lie in an earlier commit of the provider, in which case
$\Delta_{\mathrm{provider}}$ is empty. When $\Delta_m$ is part of a
collateral evolution, $(c, m)$ is a candidate task. Given the code of $m$ before $c$, a
method may modify only the files of $m$, and must make $m$ follow the
interfaces as of $c$ while preserving the existing
functionality of $m$ (Figure~\ref{fig:task-formulation}).

\paragraph{Input.}
A method receives all source files of module $m$ at the parent of $c$,
together with the kernel source tree at $c$. The tree is used for building,
and the method may also consult the support library's headers, implementation,
and documentation at $c$. The method does not see the change of $c$ to $m$,
the reference answer, the commit message or history, or the changes other
drivers made in the same commit. Depending on the evaluation setting, a method
may additionally receive compiler diagnostics obtained after reverting
$\Delta_m$, or annotated old-API names and their locations in the initial module.
Section~\ref{sec:experiments-setup} describes the settings.

\paragraph{Reference answer.}
The reference answer is $\Delta_m$, the original commit's complete change to
the module. The Linux submission rules require each patch to contain a single
logical change~\cite{kernel_submitting_patches}, and Padioleau et al.\ note
that, because this advice is largely followed, patches containing collateral
evolutions can generally be distinguished from those containing
device-specific changes~\cite{padioleau2008coccinelle}. $\Delta_m$ therefore
in general contains only module $m$'s response to the interface change,
without unrelated edits. The exceptions are five tasks whose commits also
contain unrelated edits, which we remove from the reference answer when they
occupy separate hunks (i.e., contiguous blocks of changed lines in the diff)
in 3 tasks, and keep when they share hunks with required changes in the
other 2.

\subsection{Benchmark Construction}
\label{sec:benchmark-construction}

\paragraph{Finding candidates.}
\bench is built from driver changes under \texttt{drivers/} and
\texttt{sound/} in three Linux mainline release updates: v6.19$\to$v7.0,
v7.0$\to$v7.1, and v7.1$\to$v7.2. Since an interface change induces similar
changes in the drivers that depend on
it~\cite{padioleau2006collateral,padioleau2008coccinelle}, we search these
updates for such similar changes and locate each to a commit and a module,
which gives candidate pairs $(c, m)$. This step yields 12,254 candidates in
3,872 commits.

\paragraph{Filtering.}
Interfaces are used in arbitrary code, so the forms that such changes take
cannot be enumerated in advance~\cite{padioleau2006collateral}. From these
candidates, we identify 221 preliminary groups of similar changes that appear
to share a cause. We then examine each candidate and group
manually~\cite{padioleau2008coccinelle}, using the driver code and the commit
history to confirm what caused the similar changes, and apply two conditions.

First, we exclude a candidate if its change is (1) \emph{unexplained}:
the triggering interface change, or how the driver responds to it, cannot be
established; (2) \emph{out of scope}: it lies outside driver code, such as in
framework core or tests, or it only relocates code; (3) \emph{not a shared
interface}: the interface does not come from code that provides common
services to a class of devices; (4) \emph{incidental}: the new interface is
adopted as a by-product of a bug fix, a new feature, or a driver-internal
refactoring; or (5) \emph{cosmetic}: it only touches header inclusions,
constants, tracepoint arguments, or coding style.

Second, we require that the module build at both $c$ and its parent. We
then revert $\Delta_m$ at $c$ and rebuild. If the rebuild fails, the candidate
becomes a build-breaking task. If it succeeds, we keep the candidate only when
the new usage was recommended at the time of $c$, as shown either by an
explicit statement in the kernel documentation, a code comment, or a provider
or maintainer commit that the old usage is deprecated or should be
replaced~\cite{kernel_deprecated}, or by the same replacement having already
been made by several developers in different drivers before $c$, and record
it as a build-compatible task.

\paragraph{Sampling.}
Each of the resulting 91 groups corresponds to one confirmed interface change
and is identified by the interface element that the drivers adapt to. An
interface change can affect anywhere from a few drivers to thousands. To
keep individual changes from dominating the benchmark, we sample at most five
tasks from each group, preferring tasks from different commits,
then from different subsystems and authors, and finally balancing change
size (Table~\ref{tab:benchmark-overview}).

\paragraph{Generating task inputs.}
Each task is instantiated as a method-agnostic task package: at commit $c$,
module $m$ is restored to its parent version and packaged with the build
target and a task description. A method works on this state and outputs a
patch. The evaluator applies the generated patch to a fresh copy of this task
input and reports whether the target module builds. Optional guidance is
exposed according to the experimental setting.

\begin{table*}[t]
  \caption{Overview of \bench by release update. Totals for groups and
    modules count each group or module once and are therefore smaller than
    the sums over release updates.}
  \label{tab:benchmark-overview}
  \vskip 0.15in
  \begin{center}
  \begin{small}
  \begin{tabular}{lrrrrrr}
  \toprule
  Release update & Groups & Tasks & Commits & Modules & Build-breaking & Build-compatible \\
  \midrule
  v6.19$\to$v7.0 & 33 & 101 & 74 & 94 & 43 & 58 \\
  v7.0$\to$v7.1  & 38 & 116 & 65 & 104 & 50 & 66 \\
  v7.1$\to$v7.2  & 32 & 99 & 63 & 93 & 41 & 58 \\
  \midrule
  Total & 91 & 316 & 202 & 268 & 134 & 182 \\
  \bottomrule
  \end{tabular}
  \end{small}
  \end{center}
  \vskip -0.1in
\end{table*}

\subsection{Benchmark Features}
\label{sec:benchmark-features}

\paragraph{Coverage.}
The tasks are spread across many subsystems rather than concentrated in a
few. Counting a subsystem as the first two directory levels of a module's
path, such as \texttt{drivers/net}, the 316 tasks come from 53 subsystems.
The largest, \texttt{drivers/gpu} and \texttt{drivers/net}, each account for
15\% of the tasks, followed by \texttt{sound/soc} with 7\%. 45 of the 91
groups reach the cap of five tasks.

\begin{figure}[t]
  \centering
  \includegraphics[width=\columnwidth]{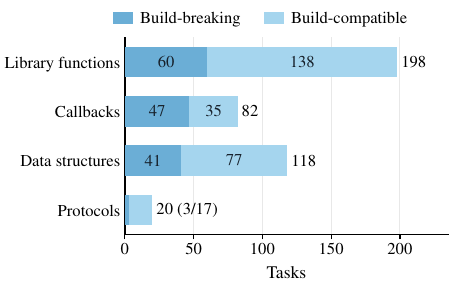}
  \caption{Tasks by interface kind, stacked by build-breaking and
    build-compatible. A task is counted under every kind it involves: 230
    tasks involve one kind, 71 two, 14 three, and 1 all four.}
  \label{fig:interface-kinds}
  \Description{Stacked bars for library functions, callbacks, data
    structures, and protocols.}
\end{figure}

\paragraph{Interface kinds.}
The tasks cover all four interface kinds
(Figure~\ref{fig:interface-kinds}). Counting each task once under every kind
it involves, 198 tasks involve library functions, 118 data structures, 82
callbacks, and 20 protocols, and 86 tasks (27\%) involve more than one kind.

\paragraph{Task size.}
The changes are small relative to the modules that contain them. The
reference answer has a median of 8 changed lines, and 62\% of the tasks
change several hunks or several files, while the target module has a median
of 1,089 lines of code. Table~\ref{tab:task-size} gives these sizes for each
task type.

\begin{table}[t]
  \caption{Task size (median / P90). Module size counts the source
    files and local headers of the Kbuild target, excluding comments and
    blank lines.}
  \label{tab:task-size}
  \vskip 0.15in
  \begin{center}
  \begin{small}
  \setlength{\tabcolsep}{4pt}
  \begin{tabular}{lcc}
  \toprule
  Metric & Build-breaking & Build-compatible \\
  \midrule
  Module lines of code      & 1,534 / 45,913 & 929 / 4,746 \\
  Module files              & 3 / 18 & 2 / 11 \\
  Reference lines changed   & 5 / 34 & 11 / 32 \\
  Reference files changed   & 1 / 2 & 1 / 1 \\
  Reference hunks           & 2 / 5 & 2 / 6 \\
  \bottomrule
  \end{tabular}
  \end{small}
  \end{center}
  \vskip -0.1in
\end{table}

%% file: appendix/driverbench/sections/pipeline.tex
\section{\sysname}
\label{sec:pipeline}

\begin{figure*}[t]
  \centering
  \includegraphics[width=\textwidth]{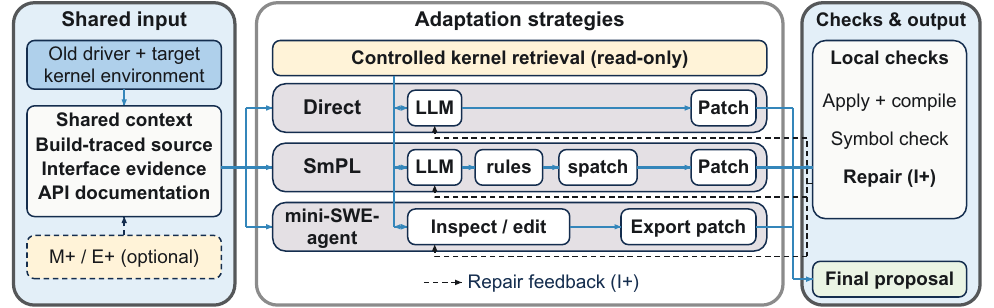}
  \caption{\sysname{} supplies shared driver and interface evidence to
    three editing strategies. Retrieval and local feedback support revision;
    the final patch undergoes independent evaluation.}
  \label{fig:agent-workflow}
  \Description{CoEvoAgent shares source and interface evidence across Direct, SmPL,
    and mini-SWE-agent. Each strategy can query permitted kernel providers
    and revise its proposal using local feedback. Optional settings add
    migration locations and initial compiler diagnostics or change the
    repair budget. Each strategy outputs a final patch.}
\end{figure*}

\sysname{} provides a common setting for studying driver adaptation in
\bench{}: methods receive the same initial driver and interface evidence,
but differ in how they produce and revise edits
(Figure~\ref{fig:agent-workflow}).

\subsection{Interface-Grounded Context}
\label{sec:pipeline-evidence}

A module can span files or share a directory with other targets. We trace
the old target's Kbuild compilation to identify its source files, local
headers, and build files, which are supplied to each method. Driver usages
and includes guide the selection of
provider definitions, interface changes, and documentation, including
recommendations where the old usage still builds.

All methods may search and read additional evidence from the same permitted
kernel providers and subsystem framework files. The post-change target
implementation, reference patch, other-driver migration examples, and
commit history are withheld.

\subsection{Adaptation Strategies}
\label{sec:pipeline-generation}

The strategies differ in how they express an adaptation. \textbf{Direct}
generates a patch against the original task source. \textbf{SmPL} generates
transformation rules that Coccinelle executes on a fresh copy of that source;
synthetic syntax examples illustrate rule construction.
\textbf{mini-SWE-agent} inspects and edits an isolated workspace, whose final
source changes form the patch.

Methods address evidenced incompatibilities and documented migrations while
preserving driver behavior. Direct and SmPL also report when no change is
needed or the evidence is insufficient.

\subsection{Feedback for Revision}
\label{sec:pipeline-checking}

Methods can use feedback on patch application, unavailable exports, and
compiler errors to retrieve more provider context and revise a proposal within
a bounded budget. SmPL also receives parsing, execution, and no-effect
diagnostics. Optional initial compiler diagnostics expose incompatibilities,
while annotated locations identify old API uses. Section~\ref{sec:experiments}
varies these inputs and repair. The final proposal is evaluated even if local
checks fail.

%% file: appendix/driverbench/sections/experiments.tex
\section{Experiments}
\label{sec:experiments}

Our evaluation asks what current methods can adapt, which guidance helps,
whether the findings persist across models, and how reference agreement
relates to migration quality. We report build-breaking
and build-compatible tasks separately: the former require recovering
compatibility, while the latter require recognizing a migration even
though the old driver still builds. This split makes migration selection
visible alongside the ability to produce compilable edits.

\begin{table*}[!t]
  \centering
  \small
  \caption{Default-configuration strategy comparison. Patch counts
    applicable, substantive candidates; the assessment columns give counts
    (percentages of N). Token use is averaged over all tasks, including those
    without API calls, and follows each provider's reporting.}
  \label{tab:patch-validation-results}
  \setlength{\tabcolsep}{4pt}
  \begin{tabular}{@{}llrrrrrr@{}}
    \toprule
    Method & Task class & N & Patch & Build & Strict & Cond. & Tokens (k) \\
    \midrule
    Direct & Build-breaking & 134 & 113 & 106 (79.1) & 55 (41.0) & 13 (9.7) & 392.8 \\
    SmPL & Build-breaking & 134 & 67 & 54 (40.3) & 19 (14.2) & 12 (9.0) & 553.1 \\
    mini-SWE-agent & Build-breaking & 134 & 110 & 108 (80.6) & 58 (43.3) & 21 (15.7) & 634.9 \\
    \midrule
    Direct & Build-compatible & 182 & 60 & 56 (30.8) & 6 (3.3) & 4 (2.2) & 418.1 \\
    SmPL & Build-compatible & 182 & 25 & 25 (13.7) & 2 (1.1) & 2 (1.1) & 679.1 \\
    mini-SWE-agent & Build-compatible & 182 & 43 & 41 (22.5) & 3 (1.6) & 2 (1.1) & 970.2 \\
    \bottomrule
  \end{tabular}
\end{table*}

\subsection{Experimental Setup}
\label{sec:experiments-setup}

We use the 316 tasks in \bench: 134 build-breaking and 182 build-compatible
tasks, spanning 91 interface-change groups and 268 modules across the three
release updates from v6.19 to v7.2. The method, guidance, and model
comparisons reuse these tasks, yielding 2,346 method--task runs.

All methods use the evidence boundary in Section~\ref{sec:pipeline}.
We vary migration-location hints (M), feedback-guided iteration (I), and
initial error diagnostics (E), using + and - to indicate whether each is
enabled. M+ supplies annotated old API names and their locations, I+ allows
revision using compiler feedback, and E+ provides the initial driver's
diagnostics. The default is \texttt{M-I+E-}.

Strategy and guidance comparisons use DeepSeek (\texttt{deepseek-flash})
with temperature zero and thinking disabled. The Direct/default model
comparison also uses Claude (\texttt{claude-sonnet-5-5}) at high reasoning
effort and GPT (\texttt{gpt-6.1-sol}) at low reasoning effort, with the same
task inputs and attempt budgets. All three use a 16,000-token output limit.
Direct and SmPL allow three candidate attempts
under I+ and one under I-, including malformed responses, plus five
retrieval-only requests. mini-SWE-agent retains its native interaction
loop, limited to 30 steps, 1,000,000 tokens, and three builds, one reserved
for final checking.

\paragraph{Validation.}
\label{sec:experiments-validation}
We apply each generated patch and the developer's reference patch separately
to the same initial driver in the target kernel. Build success requires an
applicable, substantive patch that compiles; empty and comment-only outputs
receive no credit. Reference agreement compares the resulting implementations:
identical source or normalized LLVM intermediate representation (IR)
establishes agreement, while differing IR is compared with
DiffKemp~\cite{malik2021diffkemp}. The analysis also
accounts for changes to declarations, initializers, callback tables, and
headers. We report \emph{strict} agreement for source or IR identity and
unconditional equality, and \emph{conditional} agreement when the comparison
requires assumptions about external behavior or header effects. Unresolved
comparisons receive no agreement credit. All rates use the selected tasks in
each class as the denominator, including missing, rejected, and unassessed
outputs. To interpret differences, we manually review compile-passing
DeepSeek Direct/default candidates whose source differs from the reference.

Build outcomes come from the final candidate's recorded compilation.
Reference analysis uses LLVM~19.1.7 and DiffKemp with a fixed x86
\texttt{allmodconfig}, with instrumentation and warnings-as-errors disabled.
Calls and compilation commands have a 600\,s timeout; SmPL execution,
mini-SWE-agent tasks, and DiffKemp function comparisons are limited to
120\,s, 1,200\,s, and 30\,s, respectively. Full settings and per-task
outcomes accompany the artifact.

\subsection{Results}
\label{sec:experiments-results}

\paragraph{RQ1: How do strategies handle the two task classes?}
\label{sec:experiments-strategies}

All three strategies obtain higher strict agreement on build-breaking
tasks (Table~\ref{tab:patch-validation-results}). Direct and mini-SWE-agent
are close: their build rates are 79.1\% and 80.6\%, and strict agreement
rates are 41.0\% and 43.3\%. SmPL reaches 40.3\% and 14.2\%, respectively.
Its deficit begins with producing applicable patches: 67 tasks yield a
patch, compared with 113 for Direct. The rule-based strategy therefore
produces fewer applicable adaptations in this setting.

Build-compatible tasks expose a different difficulty. Strict agreement
falls to 3.3\% for Direct and below 2\% for the other strategies. Direct
returns no change on 109 of 182 tasks. Because the unchanged driver already
builds, compiler acceptance alone would obscure these missed migrations.
The task split therefore reveals a challenge shared by the three
strategies: recognizing and completing an adaptation when an initial build
failure does not identify the need for one.

\paragraph{RQ2: Which guidance improves adaptation?}
\label{sec:experiments-localization}

We vary three sources of guidance for Direct, changing one aspect of the
default at a time (Table~\ref{tab:information-repair-results}). Migration
locations identify where to inspect, initial diagnostics expose existing
incompatibilities, and repair feedback reveals problems in a proposed
adaptation.

\begin{table}[!htbp]
  \centering
  \small
  \caption{Direct guidance variants, reported as counts (percentages).
    Each variant changes the indicated factor from the default
    \texttt{M-I+E-}.}
  \label{tab:information-repair-results}
  \setlength{\tabcolsep}{3pt}
  \begin{tabular}{@{}lrrr@{}}
    \toprule
    Setting & Build & Strict & Cond. \\
    \midrule
    \multicolumn{4}{@{}l}{\emph{Build-breaking (N = 134)}} \\
    Default & 106 (79.1) & 55 (41.0) & 13 (9.7) \\
    Locations (M+) & 117 (87.3) & 61 (45.5) & 15 (11.2) \\
    Single candidate (I-) & 89 (66.4) & 46 (34.3) & 11 (8.2) \\
    Diagnostics (E+) & 113 (84.3) & 59 (44.0) & 14 (10.4) \\
    \midrule
    \multicolumn{4}{@{}l}{\emph{Build-compatible (N = 182)}} \\
    Default & 56 (30.8) & 6 (3.3) & 4 (2.2) \\
    Locations (M+) & 70 (38.5) & 19 (10.4) & 7 (3.8) \\
    Single candidate (I-) & 39 (21.4) & 5 (2.7) & 1 (0.5) \\
    \bottomrule
  \end{tabular}
\end{table}

\paragraph{Locating the migration.}
Annotated API locations cover 132 of 134 build-breaking tasks and 153 of
182 build-compatible tasks. On build-breaking tasks, adding these locations
raises the build rate from 79.1\% to 87.3\% and strict agreement from
41.0\% to 45.5\%.
For build-compatible tasks, strict matches increase from 6 to 19.
Localization therefore helps even though it leaves replacement selection
to the model; strict agreement on build-compatible tasks remains uncommon.

\paragraph{Revising proposed edits.}
\label{sec:experiments-repair}

With localization hints and initial diagnostics absent, allowing repair
raises passing builds from 89 to 106 on build-breaking tasks and from 39
to 56 on build-compatible tasks. Strict matches rise from 46 to 55 and
from 5 to 6, respectively. The extra attempts thus improve buildability in
both classes, but add little reference agreement when the original usage
remains build-compatible.

\paragraph{Exposing initial incompatibilities.}
\label{sec:experiments-diagnostics}

Initial diagnostics help before the first proposal rather than only after
a candidate fails. With repair enabled and localization hints absent,
providing these diagnostics raises passing builds on build-breaking tasks
from 106 to 113 and strict matches from 55 to 59. The gain shows that initial
failure information remains useful even when the method can revise its own
generated candidates.

\paragraph{RQ3: How does model choice affect adaptation?}
\label{sec:experiments-models}

Compilation rates on build-breaking tasks are close across the three
models: 79.1\% for DeepSeek, 78.4\% for Claude, and 82.1\% for GPT
(Table~\ref{tab:model-comparison}). Build-compatible tasks show a larger
difference. GPT produces 140 compiling patches (76.9\%), compared with
56 for DeepSeek (30.8\%) and 43 for Claude (23.6\%). It returns no change
on five such tasks, compared with 109 and 39, respectively. Model choice
therefore substantially changes the production of compilable edits when
the initial driver already builds.

Reference agreement reveals a different comparison. On build-breaking tasks,
GPT obtains 44 strict matches, compared with 55 each for DeepSeek and Claude.
On build-compatible tasks, GPT and DeepSeek each obtain six strict matches,
and Claude obtains five. GPT's compilation counts include 28 build-compatible
patches flagged by the generation tool's file-scope check. Five GPT candidates
remain unassessed for reference agreement and are retained in the denominator.

\begin{table}[!htbp]
  \centering
  \small
  \caption{Direct/default model comparison. Assessment rates and average
    token use are calculated over all tasks in each class, including tasks
    without API calls.}
  \label{tab:model-comparison}
  \setlength{\tabcolsep}{4pt}
  \begin{tabular}{@{}lrrrr@{}}
    \toprule
    Model & Build (\%) & Strict (\%) & Cond. (\%) & Tokens (k) \\
    \midrule
    \multicolumn{5}{@{}l}{\emph{Build-breaking}} \\
    DeepSeek & 79.1 & 41.0 & 9.7 & 392.8 \\
    GPT & 82.1 & 32.8 & 18.7 & 527.9 \\
    Claude & 78.4 & 41.0 & 12.7 & 398.7 \\
    \midrule
    \multicolumn{5}{@{}l}{\emph{Build-compatible}} \\
    DeepSeek & 30.8 & 3.3 & 2.2 & 418.1 \\
    GPT & 76.9 & 3.3 & 4.4 & $\geq\!985$ \\
    Claude & 23.6 & 2.7 & 1.6 & 666.1 \\
    \bottomrule
  \end{tabular}
\end{table}

\FloatBarrier
\paragraph{RQ4: How does reference agreement relate to migration quality?}
\label{sec:experiments-manual-review}

We review all 103 compile-passing DeepSeek Direct/default candidates whose final
source differs from the reference: 84 with reference differences, 17 with
conditional agreement, and two with identical IR. The other 59 passing
candidates have source identical to the reference.

\begin{table}[!htbp]
  \centering
  \small
  \caption{Author review of compile-passing DeepSeek Direct/default candidates
    with non-identical source. Accepted denotes a complete migration with
    no introduced defect found.}
  \label{tab:manual-review}
  \setlength{\tabcolsep}{3pt}
  \begin{tabular*}{\linewidth}{@{\extracolsep{\fill}}lrrrrr@{}}
    \toprule
    Task class & N & Accepted & Buggy & Partial & Missed \\
    \midrule
    Build-breaking & 53 & 39 & 11 & 3 & 0 \\
    Build-compatible & 50 & 10 & 16 & 1 & 23 \\
    \midrule
    Total & 103 & 49 & 27 & 4 & 23 \\
    \bottomrule
  \end{tabular*}
\end{table}

Review accepts 49 candidates and finds introduced defects in 27;
four implement the migration only partly, and 23 miss it entirely
(Table~\ref{tab:manual-review}). These outcomes reinforce the task-class
contrast: 39 of 53 reviewed build-breaking candidates are accepted,
compared with 10 of 50 build-compatible candidates. In the latter class,
23 patches miss the target migration despite making compilable edits.

Reference disagreement also includes acceptable migrations. Review accepts
30 of the 84 candidates assessed as different, including 16 labeled as
valid alternative implementations. All 17 conditional agreements and both
IR-identical candidates are accepted. Reference agreement thus provides
a reproducible comparison with the developer migration, while review
distinguishes acceptable departures from introduced defects and omissions.

\FloatBarrier

%% file: appendix/driverbench/sections/discussion.tex
\section{Discussion and Limitations}
\label{sec:discussion}

Build-compatible tasks separate the production of compilable edits from
the completion of a migration. GPT produces compiling patches for 140 of
these tasks, compared with 56 for DeepSeek and 43 for Claude. Despite the
higher build count, GPT and DeepSeek each obtain six strict reference matches;
the GPT candidates have not been manually reviewed. Default Direct
generation with DeepSeek returns 109 no-change outputs that compile
but do not count as generated adaptations. Of its 60 applicable patches
on these tasks, 56 compile and six obtain strict reference agreement.
Among the 50 non-source-identical passing candidates, review accepts ten
and finds 23 that miss the target migration. Compilation therefore leaves
the central migration-selection problem unresolved in this task class.

Reference analysis is configuration-specific. Conditional agreements retain
assumptions about external dependencies or header effects, separately from
strict agreement. A concrete function or initializer difference identifies
a departure from the reference. Review accepts 30 of 84 such differences
for DeepSeek Direct/default, including 16 valid alternative implementations.
Strict reference agreement and review acceptance therefore capture
distinct aspects of adaptation quality. Targeted tests with device emulators and real hardware could
add evidence about behavior beyond this static comparison.

Constructing each reference patch requires reviewing module history and
interface evidence to separate adaptations and necessary supporting edits
from other changes. Future work could automate links among changes to
kernel and driver support library interfaces, affected driver uses, and
historical edits, then propose candidate reference patches for review.
These links could also help \sysname{} prepare task inputs
by locating relevant interface definitions, usage requirements, and driver code.
Historical driver changes would be used only to build the benchmark, not
as input to patch generation.

%% file: appendix/driverbench/sections/related_work.tex
\section{Related Work}
\label{sec:related-work}

\textbf{Linux driver evolution has motivated reusable rules, example-driven
transformations, and semantic-patch inference.} Coccinelle applies
semantic patches written in SmPL; SPINFER infers semantic patches from examples
of Linux changes and is evaluated on two datasets of 40 change sets
each~\cite{padioleau2008coccinelle,serrano2020spinfer}.
SpDiff, SYDIT, LASE, and A3 derive transformations from example changes and
apply them to other code; other studies
document Coccinelle use, tool-assisted patching, and the evolution of
faults in Linux~\cite{andersen2008spdiff,meng2011sydit,meng2013lase,lamothe2022a3,lawall2018coccinelle,koyuncu2017tools,palix2011faults}.
\bench{} evaluates module-level collateral evolutions, and \sysname{}
compares direct patches, SmPL transformations, and workspace editing under
shared task inputs.

\textbf{Linux porting tools and patch-backporting systems address specific
changes.} Prequel retrieves kernel-history commits; FixMorph, PortGPT, and
MigGPT transfer patches across versions~\cite{lawall2017prequel,shariffdeen2021fixmorph,li2025portgptautomatedbackportingusing,miggpt2025}.
\bench{} targets the adaptation of a driver module to evolving kernel
interfaces.

\textbf{Migration benchmarks span API updates, dependency updates, and
repository migrations.} CODEMENV covers API update knowledge and
migration~\cite{codemenv2025}. MigrationBench studies Java
migration~\cite{liu2026migrationbench}, and TimeMachine-bench examines
dependency updates~\cite{timemachine2026}.
Wang et al.\ show that LLMs use deprecated library APIs in Python code
completion~\cite{wang2025libevolution}.
\bench{} assesses Linux driver adaptations through compilation and reference
agreement, complemented by author review.

\textbf{Repository agents and kernel-testing systems inform execution and
validation.} SWE-bench evaluates repository issue fixes, while SWE-agent
studies agent interaction with repository tools; kGym, Live-kBench, and
SymDrive cover kernel crashes and device-free driver testing
~\cite{jimenez2024swebench,yang2024sweagent,mathai2024kgym,huang2026livekbench,renzelmann2012symdrive}.
\bench{} targets driver adaptations to evolving kernel interfaces.

%% file: appendix/driverbench/sections/conclusion.tex
\section{Conclusion}
\label{sec:conclusion}

Driver adaptation requires more than fixing build errors: some migrations
leave the old interface usable and must be identified from interface
evolution. \bench{} captures this distinction in 316 module-level tasks
grounded in Linux developer changes. It builds on our analysis of the reach of
shared kernel interfaces and the time drivers take to adopt replacements.

For Direct generation, compiler feedback improves buildability, but strict
reference agreement remains much lower when the old driver still builds.
Annotated API locations help on these tasks, highlighting the importance of
identifying where a migration applies. Manual review also shows why evaluation
requires more than a build: compiling patches include both valid alternatives
and defective or incomplete edits. In this paper, we contribute the benchmark,
a shared adaptation framework, and a validation process that reports these
forms of evidence separately. Together, they support comparisons across both
task classes and further research on migration identification, generation,
and validation.